\documentclass[a4paper,11pt]{article}
\usepackage{pos}

\def\vsR{{\scriptscriptstyle{\mathrm R}}}                
\def\vsI{{\scriptscriptstyle{\mathrm I}}}                   
\def\fir{{\scriptscriptstyle{\text{\rm IR}}}}                 
\def\fuv{{\scriptscriptstyle{\text{\rm UV}}}}              
\def\cs{c_{\scriptscriptstyle {\mathrm S}}}               
\def\eff{{\scriptscriptstyle{\text{ef}}}}                        
\def\Trs{{\mathop{\rm tr}}}		                             
\def\Trb{{\mathop{\rm Tr}}}		                             
\def\diag{\mathop{\rm diag}}                                   
\def\identity{{\mathbb I}}			                     

\newcommand{\psibar}{{\bar\psi}}             
\newcommand{\cS}{{\cal S}}                     
\newcommand{\cT}{{\cal T}}                      

\def\R{{\mathbb R}}				    
\def\C{{\mathbb C}}				    

\title{Glue Condensate, Quark Condensate and Dirac Spectral Density}

\author*[a,b,c,d]{Ivan Horv\'ath}

\affiliation[a]{Institute of Nuclear Physics, Czech Academy of Sciences,\\
  \v{R}e\v{z} near Prague 292, Husinec, Czech Republic}

\affiliation[b]{Department of Physics and Astronomy, University of Kentucky,\\
506 Library Dr, Lexington KY 40508, USA}

\affiliation[c]{Department of Physics, The George Washington University,\\
725 21stNMW, Washington DC 20052, USA}

\affiliation[d]{Institute of Physics, Slovak Academy of Sciences,\\
Dubr\'avsk\'a Cesta 9, 845 11 Bratislava, Slovakia}

\emailAdd{ihorv2@g.uky.edu}

\abstract{I derive the regularized formula for glue scalar density (gluon
condensate) in terms of Dirac spectral density [arXiv:2509.03509],
and elaborate on its uses and meaning. Particular attention is
given to understanding of what this new formula reveals about
the relation between glue and quark scalar densities, how it relates 
to IR phase, how it clarifies the distinction between anomalous and 
spontaneous ways of breaking symmetries, and what it says about 
the relation between UV and IR in QCD.}

\FullConference{The 42nd International Symposium on Lattice Field Theory (LATTICE2025)\\
2-8 November 2025\\
Tata Institute of Fundamental Research, Mumbai, India\\}

\begin{document}
\maketitle

\section{Introduction}
\label{sec:intro}

The field-theory playground for strong interactions of nature are vectorlike SU(3) 
gauge theories with fundamental quarks. This rich landscape contains 
the ``real-world'' QCD and its useful deformations, and is believed to contain 
also theories with thoroughly different particle dynamics. Such changes in theory's 
nature are reflected in the structure of its vacuum or thermal state, and it is in 
this sense that we speak about {\em phase structure} in the set ${\cal T}$ 
of these theories. 

Important role in such considerations has historically been played by quark 
and glue scalar densities, namely 
$\langle \bar{\psi}_q \psi_q \rangle, \,q \!=\! 1,\ldots, N_f$, 
where $N_f$ is the number of quark flavors, and $\langle F^2 \rangle$.
This mainly stems from their connection to chiral and scale symmetries 
which, however, are only exact in certain corners of ${\cal T}$. Although
their approximate nature in real-world QCD provided explanations 
for some basic features of strong interactions, it is unlikely that even the full 
knowledge of these ``condsensates'' over ${\cal T}$ would reveal its global
phase structure. 

Recent discovery of IR phase~\cite{Alexandru:2019gdm, Alexandru:2015fxa}
(see Ref.~\cite{Horvath:2025ypt} for a review) ushered in a new perspective on 
these issues. Indeed, the IR phase transition changes, among other things, 
a single-component system into a multi-component 
one~\cite{Alexandru:2019gdm, Alexandru:2015fxa, Meng:2023nxf} 
(IR-Bulk separation). Such change not only represents a true phase 
transformation, but it can also occur anywhere in ${\cal T}$ since its 
existence doesn't a priori rely on the standard symmetry considerations. 

Symmetry still happens to make its appearance in IR phase transition 
but in a different guise than in standard cases~\cite{Alexandru:2019gdm}. 
One possible scenario, albeit previously never seen, could be that, 
upon breaking up of a non-symmetric single-component system by the IR 
transition, one of its acquired components becomes symmetric. 
While seeming unlikely, this is in fact what happens in thermal-induced 
IR phase transition where the IR component becomes effectively scale 
invariant~\cite{Alexandru:2019gdm}.\footnote{Given that, one could use 
the language of emergent symmetries here albeit that is more typically 
used in connection with quantum phase transitions.}
We emphasize that in theories with scale invariant action at zero temperature 
(e.g. QCD with sufficiently small number of massless flavors), thermal 
IR transition represents the restoration of anomalously broken scale 
invariance with an unexpected twist: the symmetry is only restored in deep-IR 
degrees of freedom (IR component) whose existence was not even suspected, 
but not yet in the bulk.

\begin{figure}[t!]
\vskip -0.10in
\centering
\hskip 0.00in
\includegraphics[scale=0.70]{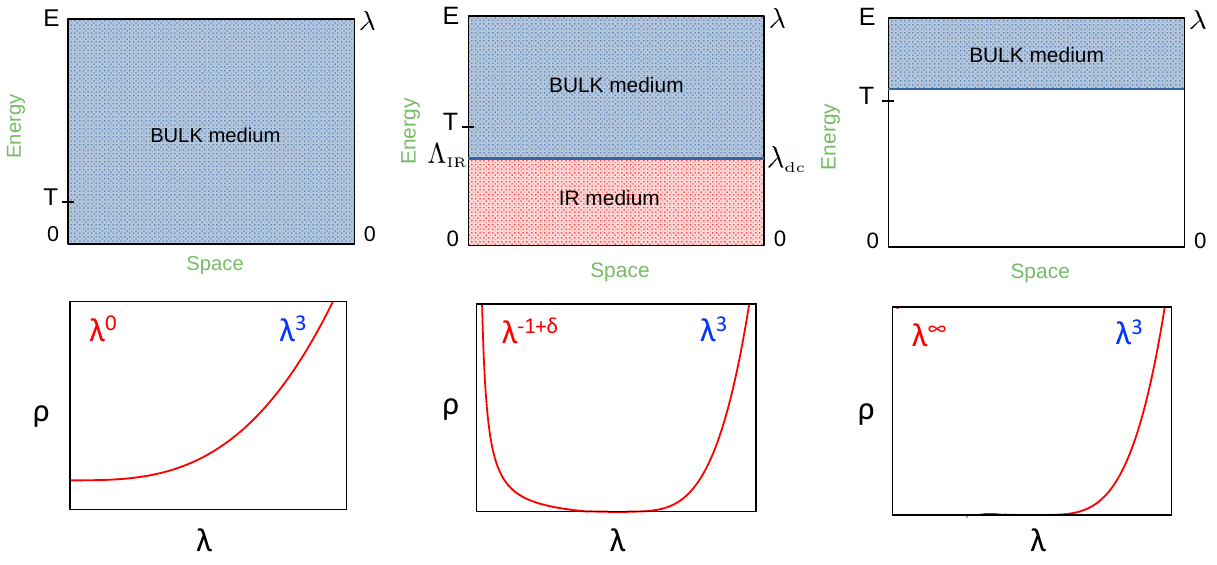}
\vskip -0.07in
\caption{Three types of thermal states (phases) in theories from 
${\cal T}$, and the behavior of associated Dirac spectral densities 
$\rho(\lambda)$. Here $\lambda$ is the Dirac spectral scale 
($ i \lambda$ is the Dirac eigenvalue in the continuum). 
Left: B phase involves a scale-broken single-component system. 
The leading IR behavior $\lambda^0$ includes cases with logarithmic 
divergence.
Middle: IR phase involves a multi-component system with IR decoupled 
from the bulk. $\Lambda_{\scriptscriptstyle{\rm IR}}$ is the energy scale 
of IR-bulk separation and $\lambda_{dc}$ the corresponding Dirac scale. 
Right: possible UV phase involves a single-component system 
of weakly interacting quarks and gluons. 
Adapted~from~Ref.~\cite{Horvath:2025jzl}.
}
\vskip -0.13in
\label{fig:IR-Bulk1}
\end{figure}

The appearance of scale invariance is connected to the fact that 
the IR-Bulk separation is actually scale-based. More precisely, it is
parametrized by the Dirac spectral variable $\lambda$ (Dirac scale) which 
thus serves as space in which the components of IR phase are separated. 
The associated distribution of quark degrees of freedom is then given 
by the spectral density $\rho(\lambda)$, and the three distinct phases in 
${\cal T}$ based on these considerations are shown and described by
Fig.~\ref{fig:IR-Bulk1} (see Refs.~\cite{Alexandru:2019gdm, Horvath:2025jzl}). 
Note the ``IR lobe'' and the ``Bulk lobe'' present in $\rho(\lambda)$,
representing components of a system in the IR phase. Scale invariance of 
the IR component is encoded by the clean inverse power law in the IR.

The above makes it clear that in order to analyze questions of scale 
invariance in the IR phase, it would be desirable to have a Dirac-scale 
decomposition of relevant observables at our disposal. More specifically, 
since violations of scale invariance due to the glue field and the quark 
field of mass $m$ are quantified by the trace anomaly, 
namely~\cite{Nielsen:1975, Collins:1977a, Nielsen:1977}
\begin{equation}
    T_{\mu\mu} \,=\, \frac{\beta(g)}{2g} \, 
    \langle F^2 \rangle  \,+\, \bigl( 1+ \gamma_m(g) \bigr) 
    \,m \langle \bar{\psi}\psi \rangle
    \label{eq:005}
\end{equation}
where $T$ is the energy-momentum tensor and $\beta$, $\gamma_m$ 
the associated RG functions, it is the decomposition of scalar densities 
$\langle F^2 \rangle$ and $m \langle \bar{\psi}\psi \rangle$ that are most 
useful in this regard. In this talk we aim to derive and analyze these
scale decompositions~\cite{Horvath:2025jzl} in a fully regularized setting.

We emphasize that, in terms of novelty, the chief benefit of this work 
(see Ref.~\cite{Horvath:2025jzl} for the full account) are the formulas 
for $\langle F^2 \rangle$ since the connection of gluon condensate 
to Dirac spectral density is counterintuitive and has not been invoked 
before. The associated thinking though is rooted in 
works~\cite{Horvath:2006az, Horvath:2006md} on coherent lattice
QCD, where all elements of the theory in ${\cal T}$ are expressed in 
terms of suitably regularized Dirac matrix, and relies 
on Ref.~\cite{Alexandru:2008fu} where necessary ingredients were put 
in place for the specific case of the overlap Dirac operator.  
  
\section{General formulas and general lessons}

Since lattice-regularized Dirac operators cannot be strictly anti-Hermitian 
due to the doubling problem, and the spectrum is thus complex, general 
expressions we are seeking naturally involve the surface spectral density
in complex plane rather than the usual line density. More precisely, with
$\lambda \!=\! \lambda_\vsR \!+\!  i\lambda_\vsI$ being the spectral 
complex variable, surface density $\rho_s(\lambda)$ involves counting 
the eigenstates contained in $d\cS \!=\! d\lambda_\vsR d\lambda_\vsI$ 
around $\lambda$. One can then show (see~\cite{Horvath:2025jzl} and 
below) that
\begin{equation}
    -m \langle \psibar\psi \rangle  = 
    m \int_{\R^2[\C]} d\cS \;  \frac{1}{\lambda+m} \;\, \rho_s^\eff(\lambda)    
    \qquad\qquad\;
    \langle \,F^2\, \rangle = \frac{a}{\cs} \, 
    \int_{\R^2[\C]} d\cS \;  \lambda \;\, \rho_s^\eff(\lambda) 
    \qquad
    \label{eq:015}      
\end{equation}
Here the effective density $\rho_s^\eff \equiv \rho_s \!-\! \rho_{s0}$ 
involves the subtraction of free field density.  Both the UV cutoff ($1/a$)
and the IR cutoff ($1/L$) are in place, and the constant $\cs$, which must
be non-zero, characterizes the lattice Dirac operator $D$ in question 
(see below). Note that in the quark case, Dirac-scale decompositions of 
this type were used in connection with the Banks-Casher 
relation~\cite{Banks:1979yr}, albeit not in this form and not at this level 
of generality.  The glue expression is entirely new.  

Equations in (\ref{eq:015}) immediately reveal some important scale-related 
features of scalar densities. Indeed, while the kinematic factor in
the quark expression enhances IR and suppresses UV, it is exactly
the opposite in the glue case. In fact, for large class of spectral densities,
the quark expression appears to become entirely IR-dominated in the chiral 
limit, and the glue expression entirely UV-dominated in the continuum limit
(note the prefactor $a$).  
However, more rigorous statements can only be made when these 
expressions are cast into the form with scale variable suitable for taking
the continuum limit (see the next section).     

\section{Glue: general formula and overlap formula}
    
Derivation of the formula for gluon condensate in \eqref{eq:015} starts with 
the following expression for the classical continuum 
limit~\cite{Horvath:2006az, Horvath:2006md, Alexandru:2008fu} 
of the lattice quantity on the left 
\begin{equation}
     \Trs_{cs} \hat{D}_{x,x}(U) - \Trs_{cs} \hat{D}_{x,x}(\identity)  \,=\,  
     \cs \,a^4 \,\Trs_c\, F_{\mu\nu}F_{\mu\nu}(x,A)  + {\cal O}(a^6)
     \quad
     \label{eq:025} 
\end{equation}
i.e. $U \!\equiv\! \{U_\mu(x)\}$ is the transcription of a classical continuum 
SU(3) gauge field $A$ onto the hypercubic lattice, and 
$F_{\mu\nu}=\partial_\mu A_\nu - \partial_\nu A_\mu + i \,[A_\mu,A_\nu]$.
The hypercubic lattice setup here consists of $N_s^3 \times N_\tau$ sites 
with UV cutoff $1/a$, IR cutoff $1/L$ ($L \!=\! N_s a$), and temperature 
$T$ ($1/T \!=\! N_\tau a$). Matrix $\hat{D}(U) \!=\! a D(U)$ is a lattice 
Dirac operator assumed to be local, gauge covariant and respecting 
hypercubic symmetries. $\identity \!\equiv\! \diag\{1,1,1\}$ denotes the free
field,  $\Trs_c$ the trace in color, and $\Trs_{cs}$ the trace in color-spin.
Eq.~(\ref{eq:025}) implies that when $\cs \!\neq \! 0$, then local 
operator $F ^2 (x) \equiv  \Trs_c \, F_{\mu\nu}F_{\mu\nu}(x)$ can be 
defined via indicated local Dirac traces which, upon utilizing translation
invariance, leads to the following expression for the quantum average    
\begin{equation}
     \langle \,F^2\, \rangle = \frac{a}{\cs} \,\frac{T}{L^3} \,
     \Bigl\langle \, \Trb \Bigl[ D(U) - D(\identity) \Bigr] \,\Bigr\rangle 
     \quad\; , \quad\;
     \langle \, \Trb \, D \,\rangle \,=\, 
     \frac{L^3}{T} \int_{\R^2[\C]} d\cS \, \lambda \; \rho_s(\lambda)  \quad
     \label{eq:030}   
\end{equation}
Here ``$\Trb$'' denotes the full Dirac trace. The second equation follows 
from the definition of surface spectral density and 
$\rho_s(\lambda) \!\equiv\!  \langle \rho_s(\lambda, U) \rangle$~\cite{Horvath:2025jzl}. 
Eq.~\eqref{eq:030} implies the formula for gluon condensate in~\eqref{eq:015}.

In order to obtain the regularized Dirac-scale decomposition of gluon condensate
in the most continuum-like setting, we evaluate it for the family of overlap Dirac 
operators $D \!=\! D(\Delta)$ based on Wilson-Dirac matrix~\cite{Neuberger:1997fp}, 
defined by
\begin{equation}
   \frac{a}{\Delta} \, D  \,=\, 
   1 + \frac{\hat{D}_W- \Delta}{\sqrt{\bigl( \hat{D}_W - \Delta \bigr)^\dagger
                                                        \bigl( \hat{D}_W - \Delta \bigr) }}
    \quad\quad , \quad\quad
    \Delta \in (0,2)
     \label{eq:035}       
\end{equation}
Here $\hat{D}_W$ is the dimensionless Wilson-Dirac operator. 
In Ref.~\cite{Alexandru:2008fu} it was shown that 
$\cs \!=\! \cs(\Delta)\!\neq\! 0$ and these operators can thus be used to 
scale-decompose the gluon condensate. The spectrum of $D$ lies 
on a circle with eigenvalues satisfying
$a(\lambda_\vsR^2 + \lambda_\vsI^2) = 2\lambda_\vsR \Delta$. 
Parametrizing this circle in polar coordinates leads to the change
of variables in \eqref{eq:015} involving 
$\rho_s(\sigma \cos\varphi, \sigma \sin\varphi)  \!=\! 
\rho(\sigma)\delta(\varphi-\cos^{-1}(a\sigma/2\Delta))/\sigma$ upon
which we obtain in terms of continuum-like variable $\sigma$ 
\begin{equation}
     \langle \,F^2\, \rangle_{a,L} \;=\; \frac{a^2}{\cs\Delta} \,
      \int_{0}^{\bigl(\frac{2\Delta}{a}\bigr)^-} \! d\sigma \, \sigma^2  \, 
      \rho^\eff(\sigma, a, L)      \;\, + \;\,\,
      T \, \frac{\langle n_0 \rangle_{a,L}}{L^3} \, \frac{2\Delta}{\cs}
      \qquad
      \label{eq:040}   
\end{equation}
Here $\langle n_0 \rangle$ is the average number of exact zeromodes equal
to the number of modes with real eigenvalue $2\Delta/a$. The latter modes 
generate the second term in \eqref{eq:040}, vanishing 
in the thermodynamic~limit.

\section{Glue: contributions from IR scales and anomalies}

Formula \eqref{eq:040} can now be used to analyze the contribution of 
different scales to gluon condensate. What can be seen immediately is
that, upon taking the thermodynamic limit, the contribution from the window
$(\sigma_1, \sigma_2)$ of finite scales is given by 
\begin{equation}
     \langle \,F^2\, \rangle_{a} \;=\; \frac{a^2}{\cs\Delta} \,
      \int_{\sigma_1}^{\sigma_2} \! d\sigma \, \sigma^2  \, 
      \rho^\eff(\sigma, a)  
      \;\; \longrightarrow \;\;  0 \quad\; \text{for} \quad\; a \,\to\, 0      
      \qquad
      \label{eq:045}   
\end{equation}
Hence, the contribution to $\langle \,F^2\, \rangle$ due to any finite range 
of scales vanishes. The argument readily goes through also for finite 
renormalized windows~\cite{Horvath:2025jzl}.  This allows us to conclude that, 
from the standpoint of Dirac scales, gluon condensate is a purely UV quantity.  
This seems somewhat surprising given the connotation the gluon condensate
acquired historically, e.g. as one of the leading terms in QCD sum 
rules~\cite{Shifman:1978bx}. 

Given the scale anomaly expression \eqref{eq:005}, an alternative way of 
conveying the above result is that the glue contribution of any IR part to scale 
anomaly vanishes, or that scale anomaly due to glue is strictly a UV effect. 
Note that, in this guise, the UV nature of gluon condensate appears much 
more natural. Indeed, $U_\text{A}(1)$ anomaly is normally associated with 
strong quantum fluctuations and the necessity to regularize the theory in UV. 
In other words, it is a UV effect. Our analysis here explicitly shows that this 
is also the case for the glue part of scale anomaly.

On the other hand, $\langle \bar{\psi} \psi \rangle$ is an IR-weighted quantity 
that also contributes to scale anomaly. Its IR composition probably has to do 
with the fact that it can also be a condensate associated with spontaneously 
broken chiral symmetry in certain corners of $\cT$.

\begin{figure}[t]
	\centering    
	\vskip -0.17in		    
	\includegraphics[width=0.96\linewidth]{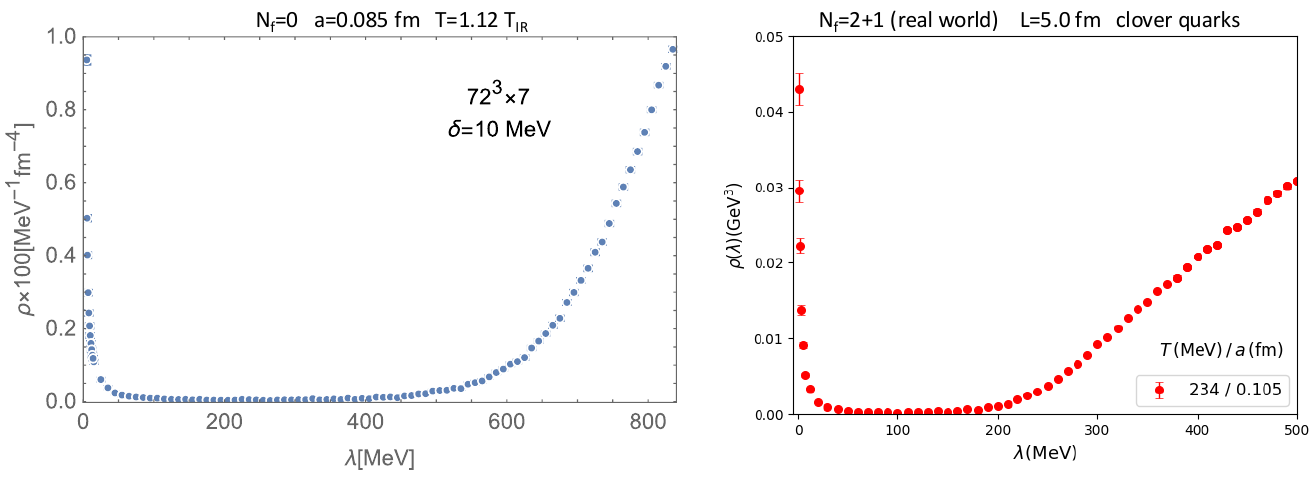}
	\vskip -0.06in	
	\caption{\label{fig:greatIR} Left: spectral density in pure-glue 
	QCD in IR phase~\cite{Alexandru:2024np}. Right: the same in 
	``real-world'',  $N_f \!= 2+1$ QCD at physical quark 
	masses~\cite{Meng:2023nxf}. They both show a pronounced 
	IR-Bulk separation.}
	\vskip -0.11in
\end{figure}

\section{Scale invariance of IR glue in IR phase.}

Among chief motivations for constructing Dirac-scale decompositions of quark 
and glue scalar densities is the scale-based multicomponent nature of systems 
in IR 
phase~\cite{Alexandru:2019gdm, Alexandru:2015fxa, Alexandru:2021pap, 
Alexandru:2021xoi, Meng:2023nxf, Alexandru:2023xho, Alexandru:2024tel}, 
and the need for analyzing scale-invariance properties of the components. 
The IR component and the Bulk are clearly separated in present-day 
simulations of thermal-induced IR phase, as illustrated by Fig.~\ref{fig:greatIR} 
for the case of pure-glue and $N_f \!= 2+1$ ``real-world'' QCD. In the latter 
case, there is a hierarchy of temperature scales
(Refs.~\cite{Alexandru:2019gdm, Meng:2023nxf, Alexandru:2024tel} are 
relevant~for~estimates~of~$T_\fir$) 
\begin{equation}
    T_c \approx \text{155 MeV}  \quad < \quad 
    T_\fir \approx \text{200-230 MeV}  \quad < \quad  
    T_\fuv \approx \text{unknown perturbative}
     \label{eq:050}       
\end{equation} 
where $T_c$ is the temperature of chiral crossover, $T_\fir < T < T_\fuv$
is the range of IR phase, and $T_\fuv$ may be infinite. In the IR phase
there exists the separation scale $\sigma_\fir = \sigma_\fir(a)$ such that
\begin{equation}
  \langle \,F^2\, \rangle = \langle F^2\, \rangle_\fir  + \langle F^2\, \rangle_B
  \quad \text{where} \quad
  \langle \,F^2\, \rangle_\fir \;=\; \frac{a^2}{\cs\Delta} \,
      \int_{0}^{\sigma_\fir} \! d\sigma \, \sigma^2  \, 
      \rho^\eff(\sigma_\fir)
      \label{eq:055}              
\end{equation}  
and the following holds
\begin{equation}
      \langle \,F^2_\fir\, F^2_{\scriptscriptstyle B} \rangle_c \;\longrightarrow\; 0   
      \quad \text{for} \quad L \to \infty   
      \qquad\;  \text{and}  \qquad\;
      \langle \,F^2\, \rangle_\fir \;\longrightarrow\; 0 
      \quad \text{for} \quad a \to 0     
      \quad    
      \label{eq:060}         
\end{equation} 
The first part expresses the fact that the IR part is in fact a component 
(its meaning is decoupling; see Ref.~\cite{Horvath:2025jzl} for more detail) 
and the second part, discussed in detail here, formally expresses its scale 
invariance. Indeed, the latter means that the contribution of IR glue medium 
to scale anomaly vanishes in the IR phase. 

\section{Where does the gluon condensate live?} 

The above discussion makes it clear that gluon condensate arises from 
the UV behavior of Dirac spectral density. Our main results, 
namely the formulas \eqref{eq:015} and \eqref{eq:040}, provide  
the necessary detail to understand how this happens. Indeed, one can 
inspect from Eq.~\eqref{eq:040}  (after taking $L \to \infty$) that 
the interplay of UV cutoff and the UV behavior of $\rho_\eff(\sigma)$ 
select out the part below,
\begin{equation}
     \rho_\eff(\sigma,a) \;=\; \frac{h(\sigma,a)}{\sigma}  \;\, + \;\, \text{rest}
     \label{eq:060}              
\end{equation}
as the relevant content of $\rho_\eff$ with respect to the gluon condensate.
Here $h(\sigma,a)$ varies slower with $\sigma$ than any non-zero power.
Indeed, higher powers lead to power divergences while the contribution
of lower powers vanishes upon removing the cutoff. Thus, the information
on gluon condensate is stored in the function $h(\sigma,a)$ similarly to
information on quark condensate being stored in deep infrared via 
Banks-Casher type~\cite{Banks:1979yr} of relationship.   

\section{Final remarks}

In this talk, we discussed Dirac-scale decompositions, introduced 
in Ref.~\cite{Horvath:2025jzl}, of quark and gluon scalar 
densities $\langle \bar{\psi} \psi \rangle$ and $\langle F^2 \rangle$.
The most original aspect of this topic is the mere existence of such 
a construct for glue. Its explicit form (Eqs.~\eqref{eq:015} and 
\eqref{eq:040}) revealed, among other things, that gluon condensate
is a strictly UV quantity. This doesn't bode well with the standard 
lore but meshes very well with the fact that $\langle F^2 \rangle$ drives 
the scale anomaly. 

\begin{figure}[t]
	\vskip -0.1in     
	\centering   
	\includegraphics[width=0.36\linewidth]{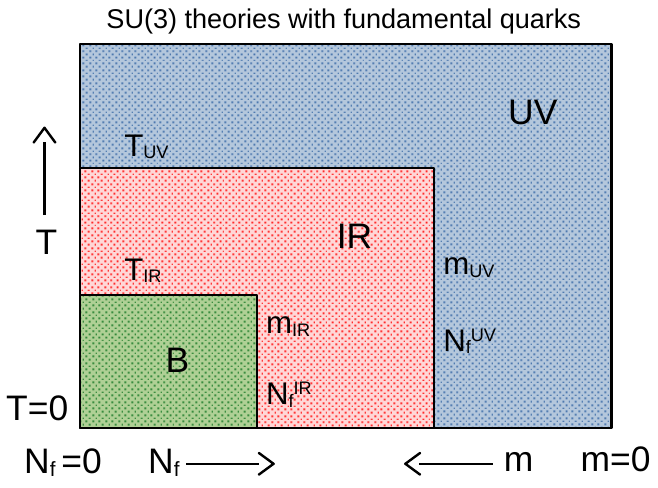}
	\hskip 0.35in
	\includegraphics[width=0.36\linewidth]{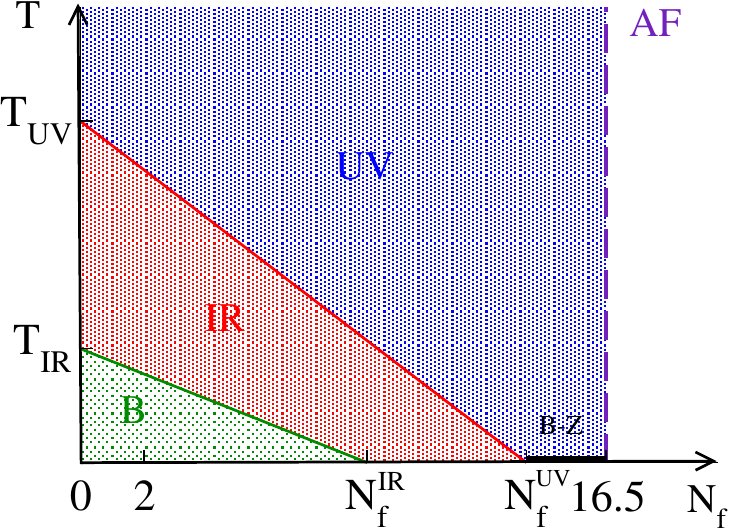}
	\caption{\label{fig:set_cT} Left: phases in set ${\cal T}$ based on 
	multicomponentality and IR scale invariance. Direction of arrows indicates 
	the direction of possible phase changes along the chain 
	$\text{B} \!\to\! \text{IR} \!\to\! \text{UV}$~\cite{Alexandru:2019gdm}.
	Right: theories with near-massless quarks. The Banks-Zaks (B-Z) regime
	and the asymptotic freedom (AF) boundary are marked.}  
	\vskip -0.1in
\end{figure}

The presented scale decomposition of $\langle F^2 \rangle$ and related 
expansions of glue observables constitute a useful tool to analyze 
the IR phase in the set $\cT$ of SU(3) gauge theories with fundamental 
quarks~\cite{Alexandru:2019gdm, Alexandru:2015fxa}.  
The phase structure derived from the existence of IR phase 
(see also Fig.~\ref{fig:IR-Bulk1}) is schematically shown in 
Fig.~\ref{fig:set_cT} for the full $\cT$ (left) and for near-massless
quarks (right). The unusual feature of IR phase is that it involves 
a multicomponent system (unlike the B an UV phases which are
single-component) and  the field-theory apparatus has not been 
previously developed for such situation. Dirac-scale decomposition
allows us to formalize the concluded IR scale invariance of the IR
component (see Eq.~\eqref{eq:060}) via its vanishing contribution 
to scale anomaly.

It is interesting to note in that regard that in the vicinity of strongly-coupled
part of conformal window $N_f \in (N_f^\fir, N_f^\fuv)$ on the right side 
of Fig.~\ref{fig:set_cT}, the corresponding zero-temperature conformal field 
theory arises from the decoupled IR component via trivial rescaling so that 
$\Lambda_\fir$ tends to infinity. The glue contribution to scale anomaly will 
then automatically vanish at every step of the procedure. 
 
\acknowledgments

This work was supported in part by the US Department of Energy, grant
DE-FG02-95ER40907.

\bibliographystyle{JHEP}
\bibliography{my-references}

%

\end{document}